\pdfoutput=1
\documentclass[10pt]{ClemsonThesis}

\usepackage{listings}
\usepackage[english]{babel}
\usepackage{csquotes}

\usepackage{biblatex}
\title{Webots.HPC: A Parallel Robotics Simulation Pipeline for Autonomous Vehicles on High Performance Computing}
\department{School of Computing}
\documentType{Honors Thesis}
\major{Computer Science}
\degree{Bachelor of Computer Science}
\graduationMonth{August}
\graduationYear{2021}
\author{Matt Franchi}
\committeeChair{Dr. Amy Apon}
\committeeMemberOne{Dr. Mashrur "Ronnie" Chowdhury}
\committeeMemberTwo{Dr. Linh Ngo}
\committeeMemberThree{Dr. Sakib Khan}
\committeeMemberFour{Dr. Ken Kennedy}
\hypersetup{
    colorlinks,
    linkcolor={black},
    citecolor={black},
    filecolor={black},
    urlcolor={black},
    pdftitle={\theTitle},
    pdfauthor={\theAuthor},
    pdfsubject={\theDocumentType},
    pdfkeywords={Clemson University, \theDepartment, \theDocumentType, \theMajor, \theDegree},
    pdfstartpage={1},
}

\begin{document}
%  ============================================================================
    \frontmatter % Begin front matter (pages are numbered with Roman numerals)
%  ============================================================================

    \addtotoc{Title Page}{\maketitle}          % Generate the title page
    \doublespacing                             % Text should be double spaced
    \setcounter{page}{2}                       % Abstract begins on page 2
    \addtotoc{Abstract}{\chapter*{Abstract}
In the rapidly evolving and maturing field of robotics, computer simulation has become an invaluable tool in the design process. Webots, a state-of-the-art robotics simulator, is often the software of choice for robotics research. Even so, Webots simulations are often run on personal and lab computers. For projects that would benefit from an aggregated output dataset from thousands of simulation runs, there is no standard recourse; this project sets out to mitigate this by developing a formalized parallel pipeline for running sequences of Webots simulations on powerful HPC resources. Such a pipeline would allow researchers to generate massive datasets from their simulations, opening the door for potential machine learning applications and decision tool development. We have developed a pipeline capable of running Webots simulations both headlessly and in GUI-enabled mode over an SSH X11 server, with simulation execution occurring remotely on HPC compute nodes. 
Additionally, simulations can be run in sequence, with a batch job being distributed across an arbitrary number of computing nodes and each node having multiple instances running in parallel. The implemented distribution and parallelization are extremely effective, with a 100\%  simulation completion rate after 12 hours of runs. Overall, this pipeline is very capable and can be used to extend existing projects or serve as a platform for new robotics simulation endeavors. }  % Generate the abstract

    %
    % The dedication page is optional.  Comment out this line if you do not
    % want to include this page.
    %
    %\addtotoc{Dedication}{\input{dedication.tex}}

    %
    % The acknowledgment page is optional.  Comment out this line if you do
    % not want to include this page.
    %
    \addtotoc{Acknowledgments}{\chapter*{Acknowledgments}
I would like to acknowledge my research partner, Rebecca Kahn, who was integral to engaging the succeeding two phases of the project.

Additionally, I would like to acknowledge all the faculty advisors of our research team. First, I recognize Dr. Amy Apon, the Tycho C. Howle Director of the Clemson University School of Computing. Dr. Apon was the faculty who first expressed interest in mentoring me for a departmental honors project. She provided invaluable experiences, advice, and feedback throughout the entire duration of the project. Dr. Apon is also responsible for bringing all of the other faculty advisors onto the project; as such, the project would simply not have been possible without her advisorship. Second, Dr. Ronnie Chowdhury and Dr. Sakib Kahn provided invaluable insights throughout the entire project; without their involvement, the team would have been lost in engaging the project's civil engineering and transportation components, which Webots.HPC was developed in mind for. Third, Dr. Linh Ngo continuously provided essential assistance in working with the Palmetto Cluster. As a past member of the Clemson CITI group and having published numerous publications involving Palmetto, Dr. Ngo devised solutions for technical challenges encountered during development, greatly facilitating development progress. Lastly, I thank Dr. Ken Kennedy for acting as an industry liaison throughout the project; his perspective was greatly influential, as it was informed by his extensive experience at BMW. Dr. Kennedy's input helped kept the pipeline from becoming too theoretical and grounded practical applications. 
}

    \singlespacing                             % Single space the lists
    \tableofcontents \clearpage                % Generate the Table of Contents

    %
    % REMEMBER: Review your caption listings in the genrated lists
    %           and make sure they include '\newline' commands as necessary.
    %           See the README for further information.
    %
    \addtotoc{List of Tables}{\listoftables}   % Generate the List of Tables
    \addtotoc{List of Figures}{\listoffigures} % Generate the List of Figures

    %
    % Include other optional lists.  Computer science, for example, would
    % likely include a 'List of Listings' (and would \usepackage{listings}
    % and \renewcommand\lstlistlistingname{List of Listings}).
    %
    %% \addtotoc{List of Listings}{\lstlistoflistings}

%  ===========================================================================
    \mainmatter % Begin main matter (pages are numbered with Arabic numerals)
%  ===========================================================================
    \doublespacing % Text should be double spaced

    %
    % Here we have each chapter in a separate file.  Name these as you choose,
    % and include them in the order you want them to appear.  Be sure to use
    % the \inputfile command.
    %
    
    \chapter{Introduction}

\section{Project Overview}
In robotics, simulation has an established tradition as a valuable tool for testing potential solutions on a virtual robot prior to engaging the overhead for real-world testing. Even so, as robots became an affordable commodity for research groups in the late 1980s and early 1990s, artificial, simulation-produce results became difficult to publish; as a result, simulation went almost out of fashion and was delegated to more niche sub-communities like swarm robotics and robot learning \cite{reckhaus_overview_2010}. However, in the last decade, simulation in robotics has once again become feasible due to rapid advances in computing capabilities. Computationally expensive algorithms can now be run on the typical personal computer, and special-purpose hardware (like HPC clusters) are now powerful enough to run thousands of instances of simulations in an affordable timespan \cite{reckhaus_overview_2010}. 

Interestingly enough, the computer gaming industry has also advanced the fidelity of robotics simulations. As the computer gaming and robotics fields share the goal of creating realistic virtual worlds, gaming's physics and rendering engines are often ported over and used as robotics simulation tools. This relationship between computer games and scientific research is more thoroughly examined in \cite{lewis_game_2002}. Overall, robotics simulations have now become extremely powerful and accurate, and thus have emerged back into the forefront of robotics research.

One of the most complex robotics applications currently in development, autonomous vehicles are set to revolutionize transportation. Presently, the American transportation complex finds itself in the largest age of transition since the invention of automobiles. The biggest automobile manufacturers are all targeting fully autonomous vehicles in the next decade; in fact, Nissan had plans to launch multiple driverless cars in 2020, last year \cite{bimbraw_autonomous_2015}. By 2040, AVs are forecasted to constitute about 50\% of vehicle sales, 30\% of vehicles, and 40\% of all vehicle travel, further bolstering the impending nature of this transition \cite{bagloee_autonomous_2016}. 

As a specific area of interest for Clemson University's Center for Connected Multimodal Mobility (C2M2), our team was tasked with developing a simulation pipeline apt for testing scenarios involving autonomous vehicles. Once developed, we would validate the pipeline through tests with a sample mixed-traffic highway merging simulation. Finally, we would take the massive output dataset created from thousands of simulation runs and extract insights through an appropriate machine learning model. 
Ignoring the directive for which this simulation pipeline would be developed, one could generalize the pipeline to a much larger set of scenarios. As Webots is separable from SUMO, any robotics simulation generated in Webots can also be run on this pipeline. The same massive amounts of output data can be produced, and equivalent insights can be extracted from said data using machine learning methods. As such, we are inadvertently tasked with developing a \textit{robotics} simulation pipeline apt for testing any scenario reproducible in Webots. 

\section{Motivation}
The primary motivation behind the development of Webots.HPC is to leverage the massive computing capabilities of High-Performance Computing (HPC) resources in running Webots robotics simulations. HPC resources can run these simulations faster and at higher throughput due to distribution and parallelization features. By introducing sources of randomization into each simulation run, an HPC-hosted sequence of one thousand or so simulation runs can produce an extremely useful large-scale dataset. Overall, porting Webots over to an HPC environment opens the doors for new applications, including real-world decision tools, robot controllers, and machine learning analysis. 

Our team's future work informed an additional source of motivation for this project. This thesis is the underlying component of an autonomous vehicle simulation project. The overall project was delineated into three phases. Phase I is this thesis: developing a pipeline capable of running Webots-SUMO simulations at scale on HPC resources (we have dubbed this pipeline \textit{Webots.HPC}) Phase II is extending a connected autonomous vehicle (CAV) highway merging simulation and running the updated simulation on Webots.HPC to produce a large output dataset. Phase III is taking the produced output dataset and extracting insights through machine learning methods. As Phase II and III were determined prior to the technical start of the project, we could preemptively accommodate for additional design needs. 

A literature review at the start (and repeated at the conclusion) of the project revealed that no published work currently exists on running Webots-SUMO traffic simulations at scale on HPC resources. Even after broadening our scope to include Webots simulations in general, we were still unable to find any previously published works. As such, to the best of our knowledge, the development work we have done so far is completely novel, which is another source of motivation.  

\section{Research Question}
The premise of this project was to create an effective pipeline for simulating robotics scenarios. As such, the research question of the project is as follows: 

Can we develop a research-grade experimental pipeline capable of running Webots and SUMO simulations at scale on HPC clusters? 

\section{Paper Overview}
This paper is structured as follows. Section 2 describes the tools and methods used in developing Webots.HPC. Section 3 provides an overview of the functionalities of Webots.HPC. Section 4 describes challenges and accompanying solutions encountered throughout the development process. Section 5 analyzes the performance of Webots.HPC, primarily focusing on the pipeline's parallelization and distribution capabilities. Section 6 contains an overview of works related to this project and also provides several avenues for future work. Lastly, Section 7 concludes the paper and reflects on the progress in answering the research question for this project.  \clearpage

    \chapter{Tools and Methods}
This section contains background information about the various components of the robotics simulation pipeline. After reading this section, one should have enough context to understand how these components fit into the workings of the pipeline. 

\section{Overview}
Webots.HPC is a pipeline, and thus owes existence to many popular tools, software, and methods. Table \ref{tab:materials} shows all of the major materials involved in the Webots.HPC pipeline development. The following subsections describe the involved project materials more at length. 

\begin{table}[]
    \centering
    \begin{tabular}{|c|c|}
        \hline
        Docker &  Containerization\\
        Singularity & Python \\
        pip & External Python Libraries \\
        SSH & X11/Xvfb \\
        Webots & SUMO \\
        Palmetto Cluster & DICE Lab Queue \\
        Portable Batch System & Distributed Computing \\
        Parallelization & Big Data \\
        \hline
    \end{tabular}
    \caption{Tools, Software, and Methods}
    \label{tab:materials}
\end{table}

\section{Docker, Singularity, and Containerization}
\subsection{Containerization}
At a broad level, Docker and Singularity are both a packaged set of platform as a service (PaaS) products that implement containerization \cite{noauthor_docker_2021}. Containerization is the packaging of software code with just the operating system (OS) libraries and dependencies that are essential to run the code; the result is a singular, low-footprint executable (called a container) that runs invariably on any infrastructure \cite{ibm_containerization_2021}. Containers have overtaken virtual machines in recent years due to their portable and resource-efficient nature.

\subsection{Docker}
As mentioned, Docker is a containerization tool. Docker is the platform that revolutionized the cloud computing field in the early 2010s, being a unified tool that allows users to easily package an application and all required dependencies into a container \cite{scheepers_virtualization_2014}. Docker is undoubtedly the de-facto standard for building containerized apps, having acquired millions of users since its inception in March 2013 \cite{noauthor_docker_2021}. Lastly, it is important to note that Docker is more popular with commercial and professional users; alternative platforms are sometimes more apt for academic usage. 

\subsection{Singularity}
Singularity is the containerization software designed explicitly for High-Performance Computing (HPC) resources (like Clemson's Palmetto Cluster) and scientific computing. Between Docker and Singularity, the end result is the same; a portable, reproducible container that can run on any environment similar to the environment the container was originally created in. A major difference between the two platforms is that Singularity's containers can be created and deployed on HPC clusters, which was a previously unmet need prior to its inception in 2015 \cite{kurtzer_singularity_2017}. Another key difference between the two platforms is that Singularity is much more secure, offering features like no privileged or trusted daemons, access to host filesystem without safety implications, a stricter privilege model, and the ability for admins to control and limit user capabilities \cite{kurtzer_singularity_2017}. 

\section{SSH, X11, and Xvfb}
\subsection{SSH}
Secure Shell (SSH) is a protocol for remote connections between computers \cite{ylonen_sshsecure_1996}. SSH is currently the most popular method for inter-computer connection, as it offers several security features like encrypted connections and password or public key-based authentication \cite{ylonen_sshsecure_1996}. As Webots.HPC is a resource for HPC environments, users typically access the pipeline by SSH-ing onto said HPC environment from their personal computer. 

\subsection{X11 and Xvfb}
X is a communication protocol that provides an abstract interface for managing a GUI. When SSH-ing into a remote computer with the intent to work with GUI-enabled applications, X11-forwarding is required for streaming the screen renderings from the remote computer \cite{nye_x_1995}. 

When running GUI-enabled applications in headless mode, you need to configure an Xvfb, or X virtual framebuffer, to send renderings to \cite{wolfe_virtual_1990}. In contrast to the X Window 11 (X11) system, Xvfb performs all graphical operations in virtual memory without showing any screen output \cite{wolfe_virtual_1990}.

\section{Python, Pip, and Dependent Libraries}

\subsection{Python}
Python is one of the most recognizable programming languages worldwide, and for a good reason. While it is a slower, interpreted language, its extreme popularity is derived from the readability of Python code, with simplified syntax made with natural language in mind. \cite{van_rossum_python_2007}

Python was factored into this pipeline from foresight, as we knew that the Webots-SUMO simulation we planned to run on the pipeline was coded almost entirely in Python. \textit{However, it is important to note that Webots also offers support for C, C++, Java, and MATLAB with identical results.}
\subsection{Pip Package Installer}
Pip is the package installer bundled with Python that allows users to install packages from the Python Package Index and other indexes \cite{ansari_prerequisites_2020}. Throughout pipeline development, it was necessary to install pip onto our singularity image, as to enable the ability to install other prerequisite libraries. 

\section{Simulation Software}
The pipeline leverages Webots as its foundational simulation software. If a simulation includes traffic simulation, the pipeline also leverages SUMO for additional support. 

\subsection{Webots}
Webots is a cutting-edge open-source robotics simulator used for robotics and transportation research. Webots offers native integration with SUMO and allows us to model an actual CAV, simulating a range of sensors found on real-world prototypes.

As Webots is the primary simulation software that our pipeline leverages, it is useful to review the way that Webots handles typical simulation tasks. Webots stores scenes with a tree structure, where the root node is the world and children nodes are the different items (robots, aesthetic scenery, and sensors, etc.) in the world. Robot nodes should always be under the root node, and also are assigned a controller. Controllers are scripts (in C, C++, Java, MATLAB, or Python) that determine a node's functionality. Controllers are also the main interface through which Robot and Sensor nodes communicate. Robot and Sensor parameters are tweaked through the Webots User Interface; however, for sensors, the sampling period (a very important parameter that influences simulation accuracy and performance) is specified in the controller for an attached robot \cite{cyberbotics_webots_2021}. 

\subsection{Simulation of Urban Mobility (SUMO)}
Simulation of Urban MObility (SUMO) is an open-source traffic simulation package with net import and demand modeling components \cite{behrisch_sumo_2011}. Since its initial development by the German Aerospace Center (DLR) in 2001, SUMO has grown into a full-featured suite of traffic modeling tools, including a road network platform capable of reading different input formats, demand generation and routing utilities from various sources (origin-destination matrices, traffic counts, etc.), a high-performance simulation usable for situations from single junctions to entire cities, and a "remote control" interface called TraCI (Traffic Control Interface) to adapt a simulation online. 

\subsection{Pairing Webots and SUMO}
Independently, SUMO is capable of simulating many different types of travel, including trains, boats, bicycles, and pedestrians, in supplement to basic vehicular travel. However, when pairing SUMO with Webots, only automotive travel is easy to simulate; this is due to a lack of pre-existing research addressing alternative transportation methods. A SUMO vehicular traffic simulation can easily be paired with Webots through the bundled SUMO Interface, which is implemented as a child node in the simulation. Most of the tunable SUMO parameters must be manually edited through the simulation's traffic configuration files (sumo.net.xml, sumo.flow.xml, sumo.edg.xml, etc.); however, opposite of sensors, the sampling period of the SUMO Interface must be specified in the Webots user interface. 

When Webots is paired with SUMO, Webots takes on the role of the user-visible simulation front-end, offering complex and customizable camera angles and highly detailed vehicle models. SUMO, on the other hand, acts as the simulation back-end, sending TraCI signals through the Webots SUMO Interface to directly control any involved traffic. As a general analogy, think of Webots as a puppet and SUMO as the puppeteer. However, Webots does also offer augmented functionality through sensors. SUMO simulations can provide extensive output, but Webots sensors can augment said output substantially. Radars, cameras, compasses, distance sensors, light sensors, and touch sensors can all be added to Webots vehicles to better understand the state of both the central vehicle and surrounding traffic. 

\section{Palmetto Cluster}
The Palmetto Cluster is a large-scale distribution testbed for scientific research applications consisting of 1,700 machines maintained by the Clemson University Cyberinfrastructure Technology Integration (CITI) group \cite{du_distributed_2018}. Palmetto is Clemson University's primary high-performance computing (HPC) resource. Through Palmetto, we are able to run multiple simulations at the same time to collect data in parallel. Within Palmetto, we utilized the Clemson University Data Intensive Computing Environments (DICE) Lab queue, which contains several high-powered, big memory-oriented computers. Table \ref{tab:dicelab_info} contains the hardware specifications of the 11 nodes that the DICE Lab has purchased.

\begin{table}[]
    \centering
    \begin{tabular}{|c|c|}
    \hline
    Make & Dell \\
    Model & R740 \\
    Chip & Intel Xeon \\
    Cores &  40 \\
    RAM &  744 GB \\
    Local Scratch & 1.8 TB \\
    Interconnect & 100g, HDR, 25GE \\
    GPUs & 2 \\
    GPU Make & Nvidia \\
    GPU Model & Tesla V100 \\
    \hline
    \end{tabular}
    \caption{DICE Lab Queue Hardware Specifications (Phase 18b)}
    \label{tab:dicelab_info}
\end{table}

The large amounts of memory and CPU cores present on these compute nodes make them ideal hosts for a pipeline that utilizes parallelization and distributed computing.

\section{Portable Batch System (PBS)}
Portable Batch System (PBS) is the software that performs job scheduling on the Palmetto Cluster, as well as the majority of other HPC clusters. The primary task of PBS is to allocate computational tasks among available computing resources. Within our project, we use the PBS job array functionality to several instances of our simulation in parallel. PBS handles all the resource allocation in the background, and end-users only have to specify what resources they need and for how long they need said resources \cite{bayucan_pbs_1998}.

\section{Distributed Computing}
In our project, most of the potential applications of this pipeline benefit immensely from a distributed computing approach. For example, in the distributed computing problem of generating a massive simulation output dataset, the dataset would be generated much faster by running two thousand simulations over fifty computing nodes versus running those same simulations on one singular computer. At the conceptual level, this is distributed computing: taking a larger problem and dividing it into many sub-tasks, each of which can be solved by one or more computers. Lastly, these computers are usually connected in some fashion, allowing for inter-computer communication. In the case of most HPC resources, this is inherent \cite{kshemkalyani_distributed_2011}. 

\section {Parallelization}
Parallelization, through parallel computing, is the simultaneous use of two or more compute resources to solve a computational problem. The following are the four steps of a parallel computing problem \cite{barney_introduction_2010}. 
\begin{itemize}
    \item There must be more than one central processing unit (CPU) present. 
    \item A problem is broken down into discrete sections that can be solved concurrently. 
    \item Each of these parts is then broken down further into a series of instructions.
    \item Instructions from each part execute simultaneously on separate CPUs. 
\end{itemize}

The parallel environment of Webots.HPC is twofold: first, at the pipeline level, batches of simulation instances are spread across that available hardware resources. Then, at the simulation level, Webots engages multi-threaded processing for each simulation instance. The effectiveness of Webots' multi-threading is evaluated in Section 5: Performance Evaluation. 

\section{Big Data}
Webots.HPC was developed with big data applications in mind. "Big" data can be defined as a dataset too large and/or too complex to be processed using traditional storage and analytic architecture \cite{lantz_potentials_2015}. HPC resources are very apt for dealing with big data due to their immense processing capabilities and specialized storage configurations. Additionally, simulation output datasets can aggregate exponentially; a simulation with a 10 MB output dataset, after being run 100,000 times in sequence, would then swell to a 1 TB size. 

\section{Chapter Conclusion} 
Overall, all of these tools, software, and methods came together to produce Webots.HPC. It is important to note that certain components are not fixed, meaning that a different commodity could have been used in its place. In Figure \ref{fig:arch_diagram}, Singularity, X11, and Xvfb are the only fixed components. The pipeline process behind Webots.HPC can be generalized further to platform another simulation software in place of the Webots and SUMO bundle. In addition, Python and Pip could be replaced with another programming language supported by this new simulation software. Even PBS is not a fixed component of the pipeline, as some HPC resources have alternative job scheduling programs installed \cite{goos_job_2001}. \clearpage

    \chapter{Pipeline Overview}
This section describes in detail the final functionality and state of the robotics simulation pipeline developed during this project. This version of the pipeline (deemed Webots.HPC) has three overarching components; the official Webots Docker image, the Singularity container created from this Docker image, and the Palmetto Cluster. As noted previously, the majority of the pipeline components are interchangeable with additional implementation overhead. Singularity is the only component deemed "fixed" in the pipeline, as this is the standard containerization platform for HPC; the reasons for this are further explained in Section II: Tools and Methods.

\section{The Finished Pipeline}
Webots.HPC is a case study, or proof-of-concept, of the pipeline developed in this project. The components of Webots.HPC were chosen due to their popularity and feasible in the fields of robotics and transportation research. While several established competitors exist, Webots and SUMO are both free-of-charge and open source, promoting them to a much larger audience of researchers, developers, and enthusiasts. In contrast, most of the viable competitors, including Paramics Modeller, Aimsun, SimTraffic, and CORSIM, are proprietary software and have thus restrictions on code modification \cite{kotusevski_review_nodate}.

\begin{figure}
    \centering
    \includegraphics{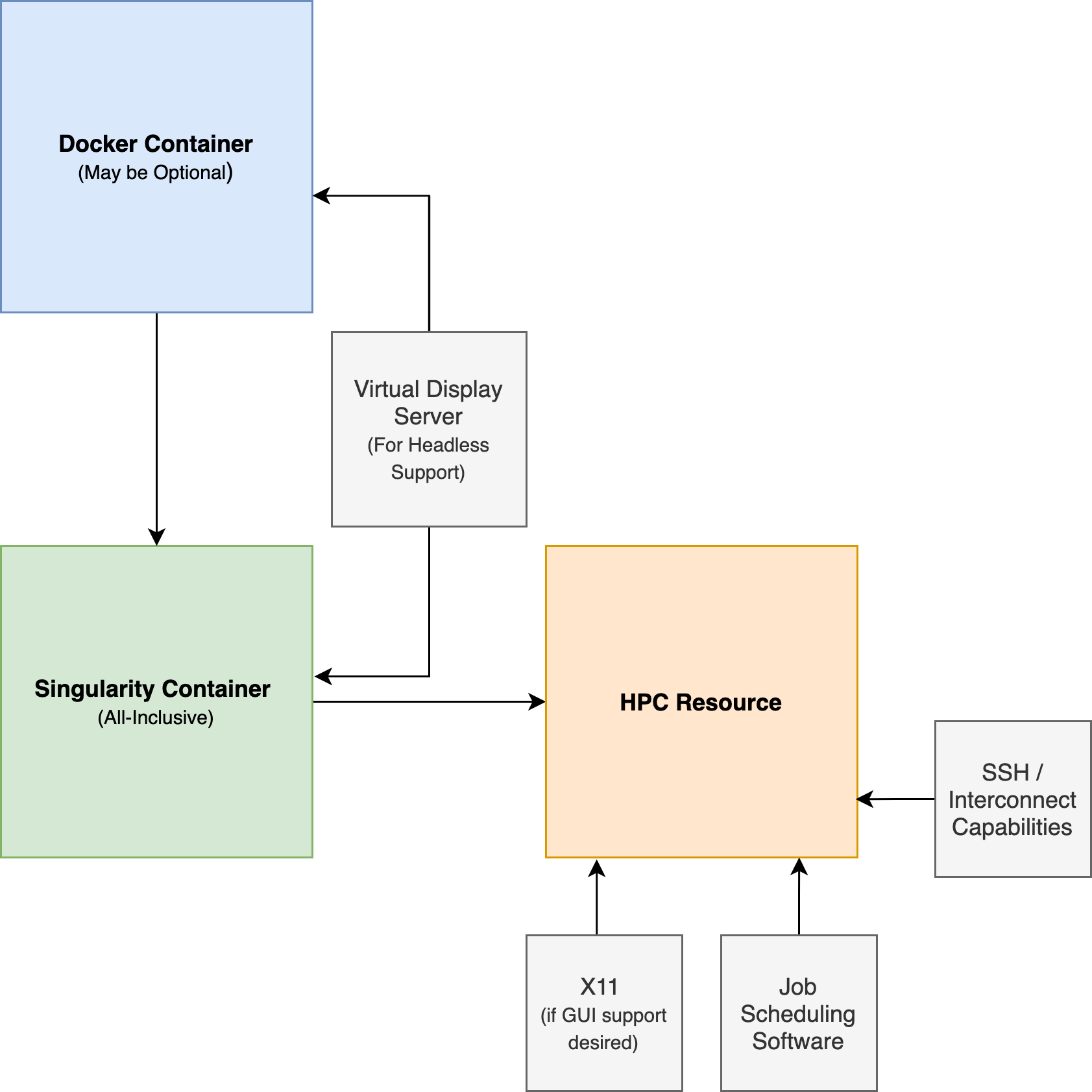}
    \caption{Underlying Pipeline Diagram}
    \label{fig:pipe_diagram}
\end{figure}

The pipeline underlying Webots.HPC is depicted in Figure \ref{fig:pipe_diagram}. Webots.HPC can be considered as a free-to-use, open-source implementation of this pipeline. Other implementations with different simulation software could have a cost overhead, as well as an entirely different set of development roadblocks and challenges. Overall, the finished pipeline is an effective way of porting simulation software meant for personal hardware over to an HPC environment; barring a direct port of their simulation software of choice, this pipeline is a researcher's best recourse for coordinating simulations on HPC. 

\subsection{Components} 
\begin{figure}
    \centering
    \includegraphics[width=\textwidth]{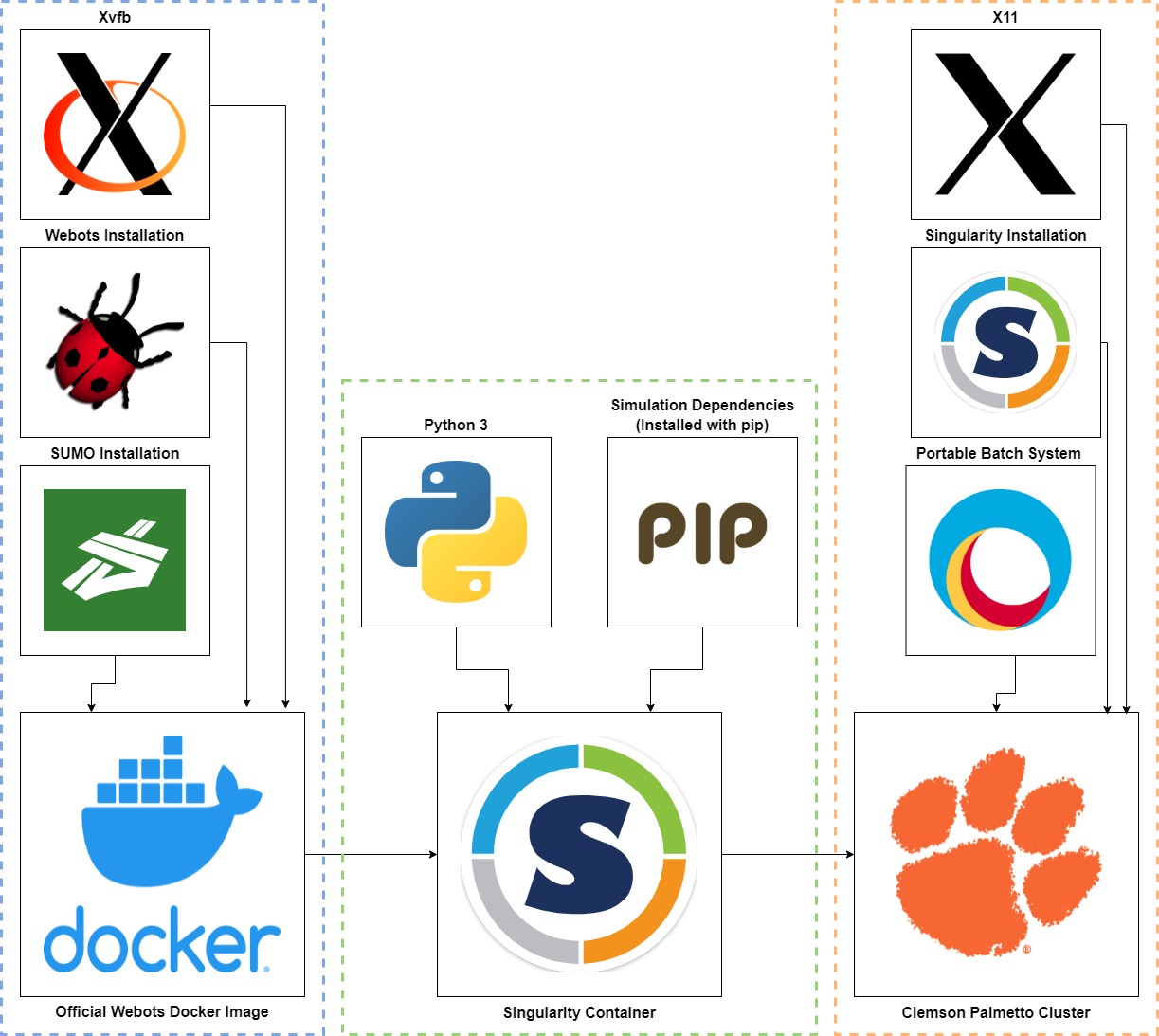}
    \caption{Architecture Diagram for the Created Robotics Simulation Pipeline}
    \label{fig:arch_diagram}
\end{figure}

The developed robotics simulation pipeline is composed of tools that are potentially useful in conjunction or independently, depending on the desired application. Figure \ref{fig:arch_diagram} provides a visualization of how the components in Webots.HPC interact. Table \ref{tab:components} lists all of the components that comprise Webots.HPC. 

\begin{table}[]
    \centering
    \begin{tabular}{|c|c|}
    \hline
    \textbf{Component} & \textbf{Purpose} \\ 
    \hline
    \textit{Official Webots Docker Image} & Commercially available, stock Webots container \\
    \hline 
    Webots Installation & Primary simulation software \\
    SUMO Installation & Secondary simulation software \\
    Xvfb Server & Virtual display server for headless mode \\ 
    \hline
    \textit{Singularity Container} & Customized, extended container for Webots.HPC \\
    \hline
    Python 3 & Chosen programming language for simulations \\
    Pip Package Installer & Official package installer for Python \\
    Pandas / NumPy & Popular data manipulation libraries \\
    Other Sample Simulation Dependencies & No explanation needed \\
    \hline
    \textit{Clemson University Palmetto Cluster} & Chosen HPC resource \\ 
    \hline
    Singularity Installation & Provides containerization for simulation software \\
    DICE Lab Queue & Big-memory hardware queue for parallel computing \\
    Portable Batch System & Job scheduling software on Palmetto \\
    X11 Server / SSH & Display server, network protocol for remote access \\
    \hline
    \end{tabular}
    \caption{Webots.HPC Components}
    \label{tab:components}
\end{table}

This robotics simulation pipeline is capable of the following functionalities: 
\begin{itemize}
    \item Running GUI-Enabled Webots Simulations on an HPC Cluster
    \item Running Webots Simulations on an HPC Cluster in Headless Mode 
    \item Running Webots Simulations that Utilize SUMO on an HPC Cluster
    \item Running Webots Simulations In Parallel on an HPC Cluster, Across \textit{n} Nodes
\end{itemize}
\subsection{Running GUI-Enabled Webots Simulations on an HPC Cluster}
If a user wants to run a simulation with a graphical component on the pipeline, they would first have to tunnel into the HPC Cluster using SSH with X11 forwarding enabled (typically, this is accomplished with the -X flag). From here, they would use the \textit{singularity exec} command to start the Webots GUI, which would automatically stream to the user's host computer. After the GUI starts, users can start their simulation identically to how they would with a local installation of Webots. 

\subsection{Running Webots Simulations on an HPC Cluster in Headless Mode}
When running a Webots simulation in headless mode, it is no longer necessary to tunnel into the HPC Cluster with X11 forwarding enabled. Instead, after SSH-ing onto a computing node of the cluster, the user would use the \textit{singularity exec} command tweaked to start Webots in headless mode, or without a user interface. See Appendix B for the exact command syntax used during the development of the pipeline. It is important to note that when starting a simulation in headless mode, you must provide a valid file path to your simulation in your command. This is because there is no GUI that users can use to manually select and start the simulation. Lastly, note that the simulation automatically starts after running the \textit{singularity exec} command. Because of this, users must build in a stop condition for their simulation, or else the Webots instance will run indefinitely. 

\subsection{Running Webots Simulations that Utilize SUMO on an HPC Cluster}
After deciding whether or not to run a simulation in headless mode, users will have to implement additional steps if their simulation utilizes SUMO. Any preprocessing of SUMO configuration files (sumo.net.xml, sumo.rou.xml, sumo.flow.xml, etc.) should be included prior to executing the \textit{singularity exec} command, as the Webots SUMO Interface only reads these files once at startup. From here, as long as the user has configured the SUMO Interface node in their Webots simulation correctly, the simulation should run as intended.

\subsection{Running Webots Simulations In Parallel on an HPC Cluster, Across \textit{n} Nodes}
Configuring a Webots simulation (with or without SUMO integration) to run in parallel on an HPC Cluster is the most involved of the four functionalities, but is consistently possible on the pipeline we developed. If a simulation does not utilize SUMO, this process is less intensive. For simulations that only use Webots, we found that adding the -a flag to the \textit{xvfb-run} command is necessary for running \textit{n} simulation instances on a compute node, where \textit{n} $>$ 1. Per the command's manual page, the -a flag instructs xvfb to try to get a free server number, starting at 99.

If a Webots simulation \textit{does} utilize SUMO, then there is an additional, somewhat menial step. To run \textit{n, where n $>$ 1} copies of a simulation on a singular compute node, we found that we had to create \textit{n} copies of the simulation on the local filesystem. These copies are all identical except for one deviation: each copy must have a unique value for the \textit{port} option on the Webots SUMO Interface node. If this value is repeated across any of the simulation copies, then some instances will fail to run, as SUMO is unable to support more than one TraCI server on the same port. During pipeline development, this process was very menial, as \textit{n} copies would have to be made each time any changes were made to the root simulation; however, in production, users could easily create a script to automate this process. It is also notable that Webots world files (.wbt files) are human-readable with any of your favorite text editors, so a script could easily be created to propagate \textit{n} copies of the simulation and then update them to have unique values for the SUMO Interface port. 

\section{Chapter Conclusion}
The pipeline developed during this project offers the ability to run simulations in any configuration of GUI-enabled mode or headlessly, and one-off or in parallel. This flexibility makes the pipeline attractive for a variety of research projects, and we hope to see some usage out of Webots.HPC in the near future. Nonetheless, our pipeline development was not without challenges, which are detailed in the following section.  \clearpage

\chapter{Challenges in Pipeline Development}
This section will describe challenges and roadblocks that we encountered during the pipeline development process. First, we will discuss in detail the causes of each problem. Then, we will present the solution to each problem. Table \ref{tab:development_challenges} describes the major challenges encountered during pipeline development.

\begin{table}[]
    \centering
    \begin{tabular}{|c|}
        \hline
        \textbf{Development Challenge} \\
        \hline
        Identifying the best method to run Webots on the Palmetto Cluster \\
        Converting the official Webots docker image to a Singularity container \\
        Modifying the Singularity container \\
        Installing additional libraries on the Singularity image \\
        Enabling GUI capabilities on the pipeline \\
        Running Webots in headless mode \\
        Enabling audio output in Webots on the Palmetto Cluster \\
        Resolving the duplicate-port issue \\
        Distributing runs across available nodes \\
        \hline
    \end{tabular}
    \caption{Pipeline Development Challenges}
    \label{tab:development_challenges}
\end{table}

\section{Achieving a Runnable State}
We faced several problems in getting one instance of a Webots simulation to run properly on the Palmetto Cluster. 
\subsection{Identifying the Best Method to Run Webots on the Palmetto Cluster}
\textit{Problem}: Before we graduated to trying to run a Webots simulation on an HPC resource, we spent a few weeks familiarizing ourselves with Webots and SUMO on our personal computers. After feeling confident enough to try and execute the same simulation on the Palmetto Cluster, we investigated the possible ways to run Webots on Linux; considering the Palmetto Cluster is Linux-based. From the Webots documentation, there are six methods of installing Webots on Linux: 

\begin{itemize}
    \item Installing the Debian Package with the Advanced Packaging Tool (APT)
    \item Installing the Debian Package Directly
    \item Installing the "tarball" Package
    \item Installing the Snap Package
    \item Installing the Docker Image
    \item Server Edition
\end{itemize}

After reviewing the available installation methods, we came to the conclusion that there was no ready-to-use distribution of Webots that would seamlessly run on the cluster. 

\textit{Solution}: After further research, we found that it was possible to convert a Docker image into a Singularity container. Knowing that Singularity was pre-installed on the Palmetto Cluster, and that Webots routinely publishes an official Docker image, we decided to try and implement this route. 

\subsection{Converting the Official Webots Docker Image to a Singularity Container}
\textit{Problem}: After identifying that the optimal route to run Webots simulations on an HPC Cluster with Singularity was through converting the official Webots Docker image, we began to actually engage the conversion process. However, we ran into permissions issues when we tried to convert the Webots Docker image on the cluster itself. 

\textit{Solution}: As modifying the permissions on the Cluster was not possible, we had to work around them instead. The solution was to pull the Webots Docker image from Docker Hub on a personal computer and \textit{then} convert the modified image into a Singularity container using the \textit{singularity build} command. 

\subsection{Modifying the Singularity Container}
\textit{Problem}: In a similar sense, once a Singularity container is on the Palmetto Cluster, it is immutable, at least at our access level. This was a problem when it came to needing to modify the container. 

\textit{Solution}: As modifying the permissions on the Cluster was not possible, we had to work around them instead. The solution was to take the Docker image that we pulled from Docker Hub on a personal computer and modify it on that same computer. After we were satisfied with the changes made to the Docker image, we would then convert it into a Singularity container using the \textit{singularity build} command. The one downside to this solution is that this process needed to be done again in its entirety each time a further change was needed. 

So, to propagate a change to the Singularity image on Palmetto, we would need to (1) pull the latest version of the Docker image to a computer on which we had admin rights, (2) make any changes locally on the image, (3) push the updated image back to Docker Hub, and (4) pull and convert the Docker image to a Singularity container on Palmetto using the \textit{singularity build} command. 

\subsection{Installing Additional Libraries on the Singularity Image}
\textit{Problem}: Other than making root filesystem changes to the Singularity container, we found that it was impossible to install additional Python libraries onto the image. We tried to solve this problem in several ways. 

\textit{Solution}: First, we created an alternate version of the Singularity container with the sandbox mode enabled. In sandbox mode, users are able to write and save changes to the typically immutable image. We believe that this approach would have worked if not for the fact that pip was not pre-installed on the image. 

We tried to install pip using the official method for Python3 but were unable to do so, receiving an 'unable to locate package' error. The typical resolution for this is to run the \lstinline!sudo apt-get update! command, but we were unsuccessful in running the command in sudo mode due to permissions limitations. As such, we decided to try another route. 

After realizing it was impossible to install pip on the converted Singularity container, we decided to move back a step and try to install it on the Webots Docker image. We were surprised that pip was not pre-installed on the Webots Docker image; however, we now believe that this was a conscious design decision for security reasons. Even so, we thought it advantageous to load our container with pip and several of the most popular Python libraries (NumPy, Pandas, etc.) to make it a better host for a large range of robotics simulations. 

We found that, on a local computer on which we had admin rights, it was possible to install pip on the Webots Docker image using the official installation script (\lstinline!python get-pip.py!). After installing pip, we were then able to load the Docker image with the aforementioned additional libraries. After we were completely satisfied with the present libraries and capabilities of the Docker image, we did a final conversion back into a Singularity container. 

\subsection{Enabling GUI Capabilities on the Pipeline}
\textit{Problem}: We first wanted to achieve running the GUI-enabled version of Webots on the Palmetto Cluster, as debugging is easier when visual inspection is possible. However, we ran into several issues with X11 on macOS and Singularity along the way. 

\textit{Solution}: While Webots' default tendency, even on an HPC resource, is to start in its GUI-enabled mode, we had to properly configure 

\subsection{Running Webots in Headless Mode}
\textit{Problem}: After we achieved running the GUI-enabled version of Webots on the Palmetto Cluster, we then graduated to trying to achieve a headless mode. In headless mode, Webots does not trigger any sort of user interface; this mode is ideal for running large-scale simulations. 

\textit{Solution}: To run Webots in headless mode on the Palmetto Cluster, we leveraged the Xvfb client bundled with the Webots Official Docker image. This Xvfb client luckily transferred over seamlessly to our created Singularity container, so we were still able to use it.

\subsection{Enabling Audio Output in Webots on the Palmetto Cluster} 
\textit{Problem}: When running Webots simulations on the Palmetto Cluster, we received audio driver errors that prevented the initialization of the Webots audio system. 

\textit{Solution}: Currently, we have no solution for this. While auditory output may be necessary for certain robotics simulations, we deemed such output as out-of-scope for this particular project. A future work would be to identify the root of these audio driver errors and enable auditory output.

\section{Running Simulations At Scale}
\subsection{Resolving the Duplicate-Port Issue}
\textit{Problem}: When running more than one Webots-SUMO simulation in parallel on a single computing node, any instances after the first would crash upon start. Upon investigation, we realized this was because SUMO was attempting to start its TraCI server on the same port for all simulations, which is a restricted behavior. 

\textit{Solution}: To resolve this duplicate-port issue, we created \textit{n} copies of our simulation, each with a unique value for the SUMO Interface port. This resolved the issue. We tended to increment the default port value of 8873 by 7 for each successive parallel simulation and ran into no further issues on this front. 

\subsection{Distributing Runs Across Available Nodes}
\textit{Problem}: During pipeline development, we identified the need for a way to evenly distribute a load of simulations across the available hardware. 

\textit{Solution}: Knowing the layout of the Palmetto Cluster, we found that using a PBS job array was the most straightforward method of job distribution. By using a job array, PBS allocates jobs across the specified hardware extremely effectively; in fact, the PBS algorithms are likely much more effective than any homegrown algorithm we could have developed during the project. 

\section{Chapter Conclusion} 
This section was included to inform readers of development challenges and provide solutions that can assist in other, similar projects by eliminating debugging time. Frustrated by the vagueness of some academic works in describing technical solutions, we tried to be as verbose as possible. Aside from this, after overcoming these challenges in pipeline development, we had achieved a polished product. For further analysis on the polish and performance of Webots.HPC, the following section evaluates the pipeline implementation and provides a performance analysis. 

 \clearpage

    \chapter{Performance Evaluation}
This section seeks to provide a concrete evaluation of the performance of Webots.HPC. To provide a performance evaluation, we asked the following questions: 

\begin{itemize}
    \item How effective is the applied distribution? 
    \item How well are the simulation instances distributed? 
    \item How effective is the applied parallelization? 
\end{itemize}

We assert that these three questions best evaluate the capabilities of the Webots.HPC pipeline, as the parallelization and distribution functionalities are the most cutting-edge and meaningful contributions to the fields of robotics and transportation research. 

%\section{File I/O}

\section{How effective is the applied distribution?}
We found that the effectiveness of the pipeline's distribution capabilities is best evaluated by measuring the speedup and throughput between running a sample simulation on a personal computer, and then on an HPC resource. On a personal computer, it is safe to say that distributed computing is not possible, as there is only one hardware unit to leverage. In our experimental test, we ran the sample simulation across six compute nodes, where each node was capable of running eight simulation instances simultaneously. Note that each compute node had specifications identical to those in Table \ref{tab:dicelab_info}, as the compute nodes belonged to the DICE Lab queue. 

\begin{figure}
    \centering
    \includegraphics[width=0.7\linewidth]{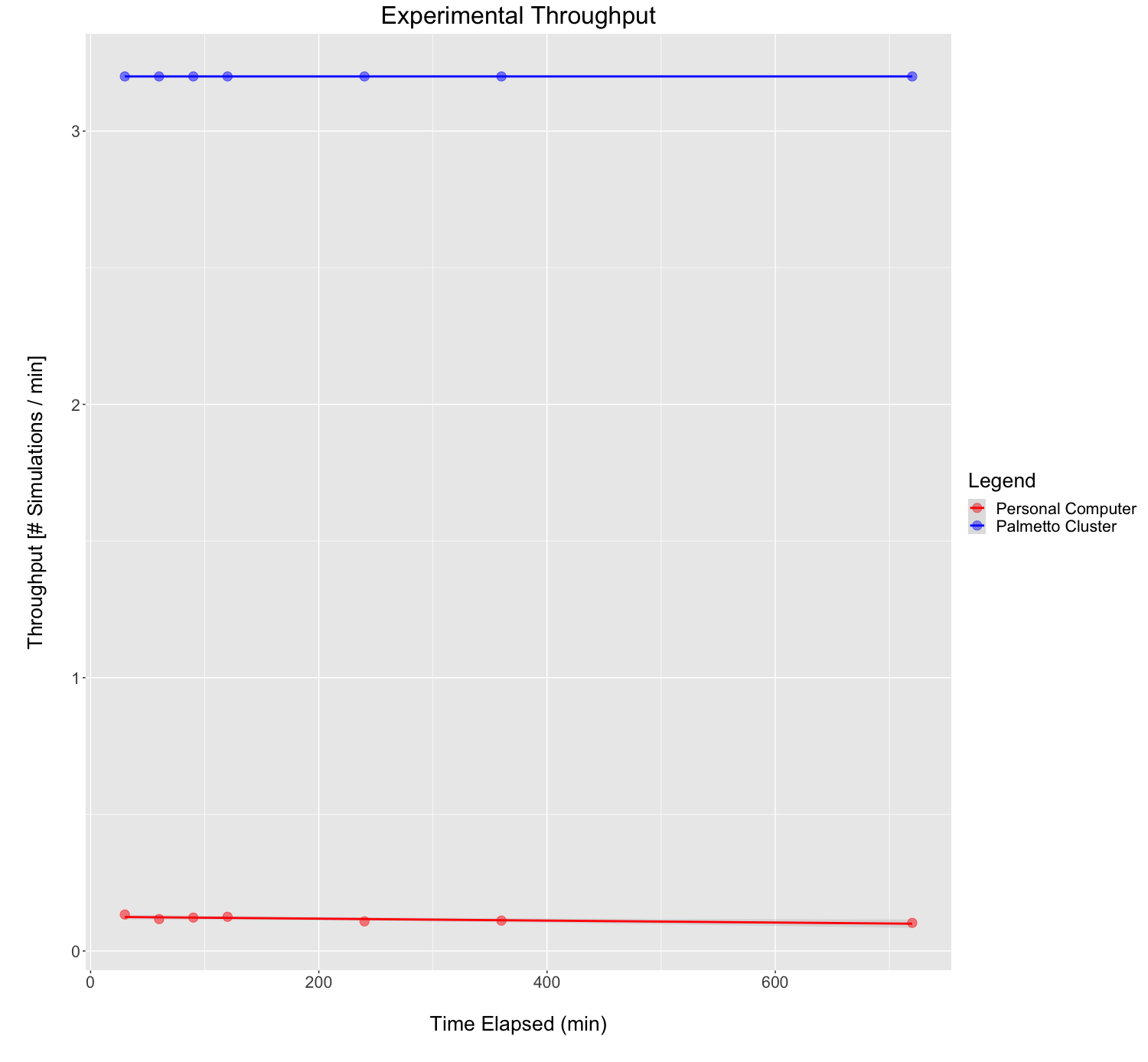}
    \caption{Sample Simulation Throughput}
    \label{fig:throughput}
\end{figure}

After our experimental test, we found that after 12 hours, the Palmetto Cluster (with compute nodes split into 8 sections) had run approximately 31 times more simulation instances than a personal computer of comparable hardware, or 2,304 runs to 74 runs. This result clearly indicates that the applied distribution is extremely effective; the PBS job scheduler present on Palmetto coordinated an even split of each group of simulations, fully leveraging the computing capabilities of each node. Figure \ref{fig:throughput} visualizes theses results. 

Lastly, it is significant to note that these results should scale with larger amounts of allocated compute nodes. For example, if we repeated this same experiment with 12 compute nodes, rather than 6, we would expect Palmetto to output approximately \textit{62} times more simulation instances, or about 4,608 runs to 74 runs. 

\section{How well are the simulation instances distributed?}
\begin{table}[]
    \centering
    \begin{tabular}{|c|c|c|}
    \hline
        Timestamp &  Personal Computer & Palmetto Cluster \\
        \hline
        30 & 4 & 96 \\
        60 & 7 & 192 \\
        90 & 11 & 288 \\
        120 & 15 & 384 \\
        240 & 26 & 768 \\
        360 & 40 & 1152 \\
        720 & 74 & 2304 \\
        \hline
    \end{tabular}
    \caption{Sample Simulation Throughput on Personal Computer vs. Palmetto Cluster}
    \label{tab:throughput}
\end{table}

Table \ref{tab:throughput} illustrates the near-perfect instance distribution that our pipeline coordinates using PBS. At each sampled timestep, the Palmetto Cluster had produced \textit{48 * t} output datasets, where t is equivalent to \textit{timestamp / 15 minutes}; t derives to this because the pipeline implemented a 15-minute walltime for each triggered job, where each job contains 48 instances of our sample simulation. This walltime is specific to the simulation running on the pipeline and will thus need to be determined prior to running a large sequence on the pipeline. Overall, PBS is capable of distributing simulation instances extremely well, allocating the correct number of simulations to each compute node (in this case, eight simulation instances to each of six compute nodes) 100\% of the time during the experiment. 

\section{How effective is the applied parallelization?}
The effectiveness of our pipeline's parallelization is partially equivalent to the performance of Webots' and SUMOs' multi-threading capabilities. As mentioned previously, parallelization in our pipeline occurs within each computing node, or within the distributed computing component. However, we did not implement any custom parallelization algorithms at the program level during pipeline development; we will merely identify how Webots and SUMO implement parallelization, as these software form the backbone of the pipeline itself. 

With regards to Webots, there are several important notes regarding parallelization. First, controller multi-threading is out-of-scope for our work with the pipeline, as users must take special care to develop controllers that utilize multi-threading; in other words, Webots does not have any native mechanism for configuring multi-threading on robots controllers \cite{cyberbotics_webots_2021}. 

Second, there \textit{are} native multi-threading capabilities for the Webots physics engine, \textit{but} users must tweak both the program level and simulation level preferences to effectively utilize these capabilities. At the program-level, users must edit the 'Number of Threads' option in the Webots preferences to match the hardware specifications of their machine or HPC resource. Then, at the simulation-level, users will need to edit the WorldInfo node and change the 'Optimal Thread Count' field to a value that is roughly half the value of 'Number of Threads'. From the Webots documentation, it is important to note that for simulations involving several robots physically independent from one another, setting a 'Optimal Thread Count' value greater than 1 can significantly improve the speed of simulation. In other cases, it may reduce the simulation speed due to the overhead of the Webots multi-threading. 

The parallelization of simulation runs is also a component of the pipeline, and we will now analyze how effective said parallelization can be. Using a sample Webots-SUMO highway merging simulation, we triggered a sequence of simulation runs with two different distributed configurations: using six nodes and running eight simulations simultaneously per node (a parallel configuration), and using six nodes each with one simulation running at a time (a serial configuration). The below table provides the theoretical computing power allotted to each simulation for each setup. 

\begin{table}[]
    \centering
    \begin{tabular}{|c|c|c|}
    \hline
    \textbf{Setup} & \textbf{6x1} & \textbf{6x8} \\
    \hline
    Cores &  40 & 5 \\
    RAM &  744 GB & 93 GB \\
    Local Scratch & 1.8 TB & 225 GB \\
    Interconnect & HDR & HDR \\
    \hline
    \end{tabular}
    \caption{Hardware Specifications for Each Experimental Setup}
    \label{tab:parallel_compare}
\end{table}

We found that the 6x8 setup had a sizably higher throughput than the 6x1 setup, even though the simulation running time was roughly the same. In fact, for most simulations, the only way an \textit{n}x1 setup would have a higher throughput would be if the splitting of computing resources was a substantial bottleneck.

Table \ref{tab:resource_consump} presents the average amount of computing resources consumed per simulation run for each experimental setup. 

\begin{table}[]
    \centering
    \begin{tabular}{|c|c|c|}
    \hline
    \textbf{Attribute} & \textbf{6x1 Setup} & \textbf{6x8 Setup} \\
    \hline
    Cores & 40 & 5 \\
    RAM [GB] & 748 & 93 \\
    Walltime & 163 & 245 \\
    CPU Time & 720 & 690 \\
    RAM Used [GB] & 2.2 & 2.3 \\
    CPU \% & 215 & 177 \\
    \hline
    \end{tabular}
    \caption{Simulation Resource Consumption Across Two Experimental Setups}
    \label{tab:resource_consump}

\end{table}

As evidenced from table \ref{tab:resource_consump}, there is not a sizable difference in performance between an \textit{n}x1 simulation configuration and an \textit{n}x\textit{\# Computers} configuration. We hypothesized that the \textit{n}x1 would have a lower CPU usage time, larger RAM usage, larger CPU percent usage, and shorter walltime. On average, the \textit{n}x1 setup has a 33.5\% shorter walltime, which is substantial, but not substantial enough to justify a serial run configuration. 

With regards to other results, the  \textit{n}x1 setup actually took 4\% longer CPU time than the parallel configuration, which was unexpected; however, we predict that this difference is insignificant and actually indicates poor native multi-threading capabilities in Webots. From the Webots documentation, controllers have to be manually configured for multi-threading, which is an intensive process \cite{cyberbotics_webots_2021}. Lastly, the \textit{n}x1 configuration uses 4.4\% less RAM than the parallel configuration, which meant one of two things: one, that our sample simulation simply uses around 2.3 GB of RAM, or Webots is unable to scale past this amount of memory (which is unlikely, due to the integrity of the software). In a future work, one could test the memory scalability of the Webots.HPC pipeline more accurately by running a simulation with a known large memory footprint. 

\begin{figure}
    \centering
    \includegraphics[width=0.7\linewidth]{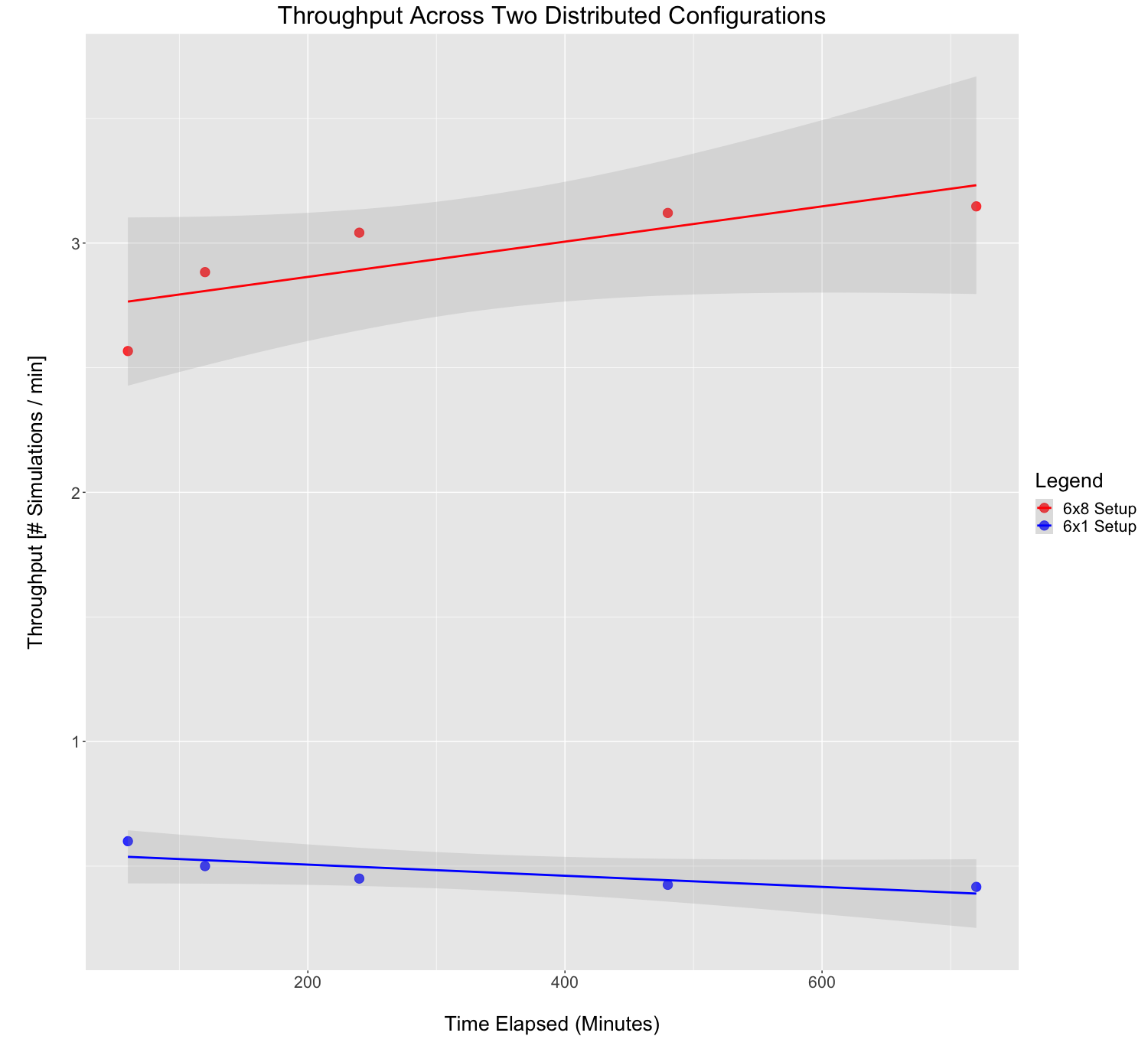}
    \caption{Parallelization Performance Across Two Experimental Setups}
    \label{fig:parallel_compare}
\end{figure}

Overall, for this particular sample simulation, it is easy to identify that a parallel configuration will achieve a much larger throughput (as evidenced by Figure \ref{fig:parallel_compare}). Even so, it would likely be helpful to test both a parallel and serial setup for a new simulation and then determine which configuration to use on a case-by-case basis. If a simulation has an extremely large memory footprint (in the range of something typically unavailable on personal hardware, like several hundred gigabytes), a serial run configuration might achieve higher throughput. However, for our sample situation, delineating powerful computing nodes into sections with specifications reminiscent of a personal computer is much more effective. 

\section{Chapter Conclusion} 
Overall, the parallelization and distribution capabilities of the Webots.HPC pipeline are effective, customizable, and powerful. Even so, there is further room for evaluation. One limitation to this project (due to time constraints) is a lack of analysis on file input and output (I/O) speeds. While we presume that the file I/O of Webots.HPC is sufficient due to the established integrity of the Palmetto Cluster and the technical reputation of Python3 and Webots, future work could include identifying exact file I/O metrics. Additionally, work could be engaged to improve file I/O rates if the determined metrics were sub-par. Even with this limitation, we are satisfied with the performance of the Webots.HPC pipeline. The following section provides a further discussion of results and provides conclusions, connections, related works, and additional avenues for future work. 
 \clearpage

    %\inputfile{relatedWork.tex}
    
    \chapter{Conclusions and Discussion}
This section first describes the preceding publications and related works that led to the development of this pipeline. Then, we provide several avenues for future work to further bolster the functionality, portability, and feasibility of the Webots.HPC pipeline. Next, we reflect on how well we answered our project research question: \textit{Can we develop a research-grade experimental pipeline capable of running Webots and SUMO simulations at scale on HPC clusters? } Lastly, we provide additional connections and closing remarks. 

\section{Related Works}
To establish the validity of this research at the broadest level, there are several papers \cite{hohl_aibo_2006}, \cite{magyar_developing_2003}, \cite{zlajpah_simulation_2008} that endorse the validity of using Webots as a platform for state-of-the-art robotics simulation. More recently, there are numerous notable publications \cite{khan_situation-aware_2020}, \cite{kornyei_multi-scale_2021}, \cite{ma_projective_2019}, \cite{karoui_performance_2017} that establish the feasibility of using Webots as a platform for autonomous vehicle simulation research. When searching for works that address the parallelization of Webots simulations and running at scale, we came up largely short. One publication, however, was promising. In 2020, a research team in Greece developed a Webots-based deep reinforcement learning (DRL) framework for robotics, dubbed \textit{Deepbots}. This project combined Webots with the open-source OpenAI Gym framework for DRL, creating a standardized interface to employ DRL across various robots scenarios. However, this project lacks any custom parallelization components and does not mention the usage of HPC resources. Besides this, the overall objective of the project is sizably different from our directive. Even so, the Deepbots project is similar in its objective of providing a standardized process for triggering simulations.

As such, to the best of our knowledge, this work is the first to address the running of Webots robotics simulations at scale, leveraging the hardware capabilities of an HPC cluster. 

We derived insights from numerous papers that essentially implemented our project with a different simulation software and different field. \cite{kornyei_multi-scale_2021} analyzes a problem within the field of environmental science, air pollution, by conducting simulations powered by SUMO. These simulations are configured to be multi-scale and leverage HPC architecture.

\section{Future Works}
\subsection{Formalizing a User Interface}
One major avenue for future work based on this project would be to make the developed pipeline more cohesive and user-friendly. This would be accomplished through a mixture of automation, user interface development, and usability testing. Several parameters, like the name of the job, the number of instances, the job queue, and the hardware requirements of the PBS script could be inputted into a user interface, rather than the current process of manually editing the script. Additionally, the job status and other reporting metrics could be triggered automatically, rather than executed manually. In sum, there is a large potential to morph this pipeline from its current state of various independent tools into a singular, easily tweakable application. 

\subsection{Scalability Testing}
Another avenue for future work would be to evaluate the scalability of our pipeline. Currently, the pipeline has only been tested on Clemson University's Palmetto Cluster, as this was the resource on which the pipeline was developed. While we predict that the pipeline is largely scalable because of hardware-level similarities across American HPC clusters \cite{martinez_validating_2008}, validating this hypothesis would be substantial enough to be a standalone future work. 

\subsection{Converting to the Cloud}
An additional avenue for future work would be to take the pipeline from this project and convert it for the cloud. Since one can make the argument that HPC clusters are private clouds \cite{mateescu_hybrid_2011}, achieving this pipeline in the cloud could enable even larger big data applications. Additionally, a cloud-enabled version of the pipeline could take advantage of several standard offerings from the major cloud providers. For example, an implementation on Amazon Web Services (AWS) could easily take advantage of autoscaling, eliminating the need for static provisioning of resources through a PBS script \cite{aws_aws_2021}. 

\subsection{A Platform for Future Projects}
Next, one can not ignore the potential of the pipeline developed in this project. Almost any Webots simulation can be easily ported to this pipeline and then harness the power of parallelization. With parallelization comes high speed and throughput, which then produces larger and larger output datasets. The latter two phases of our project do this very thing, taking a Connected Autonomous Vehicle (CAV) highway merging simulation and running it on the developed pipeline to produced merged datasets of several thousand runs. From here, such datasets can be used as inputs for real-world decision tools; or, these large sums of data can have useful insights extracted from them using the latest machine learning models.

\section{Answering the Research Questions}
In answering the research question, \textit{Can we develop a research-grade experimental pipeline capable of running Webots and SUMO simulations at scale on HPC clusters? }, we can confidently say that the pipeline we set out to develop did come to fruition. We first configured Webots and SUMO to run independently on our HPC resource, the Palmetto Cluster. After establishing a process for executing these applications on the cluster, we then moved to running Webots and SUMO in conjunction, as to enable support for vehicular simulations. After achieving this, we wrote a PBS Job Script, with tunable parameters, for triggering such a simulation either in GUI-enabled mode or headlessly. After succeeding here, we had answered the first portion of our research question (a research-grade experimental pipeline capable of running Webots and SUMO simulations) with a definitive yes. 

After achieving a single triggered simulation run on our pipeline, we moved towards the goal of running simulations at scale. By extending the PBS Job Script with job array functionality, we were able to run several instances of a simulation in parallel, with eight instances each on six high-power compute nodes. Thus, at the conclusion of the project, we can answer the entire research question affirmatively.

\section{Connections and Conclusions}
At the conclusion of our project, we provide a list of accomplishments achieved during development. In chronological order: 

\begin{itemize}
    \item We extended the official Webots Docker image with pip and other popular Python libraries.
    \item We converted said Docker image into a Singularity container, for usage on the Palmetto Cluster.
    \item We successfully used the Singularity container to run Webots in GUI-enabled mode on Palmetto. 
    \item We successfully used the Singularity container to run SUMO in GUI-enabled mode on Palmetto. 
    \item We paired Webots and SUMO together to run a sample simulation in GUI-enabled mode on Palmetto. 
    \item We converted the Webots-SUMO software to run headlessly and ran this same sample simulation again on Palmetto. Keep in mind, at this point, all simulation runs were one-off and manually triggered. 
    \item We added the ability to trigger a serial sequence of simulations through a PBS job script. 
    \item We added an additional layer to the job sequence, adding parallelization capabilities to the aforementioned script through a PBS job array. 
    \item We successfully evaluated the distribution and parallelization algorithms of the Webots.HPC pipeline. 
\end{itemize}

In summary, we feel that a substantial contribution has been made to the transportation and robotics simulation fields, primarily due to the novelty of our work. As mentioned, to the best of our knowledge, running Webots and SUMO in an HPC resource has not been previously engaged. More so, we are not only running Webots-SUMO simulations on HPC, but also harnessing the additional computing power by enabling parallel simulation sequences. These sequences open the doors for a plethora of unnamed future work; essentially, any simulation that has meaningful deviations between runs can be run in a parallel sequence to produce a dataset of any size. The high theoretical ceiling on this dataset's size makes previously infeasible applications like real-world decision tools, deep learning, and machine learning models possible, and we hope to see such applications come to fruition through the usage of our pipeline. 

Not only is the Webots.HPC implementation of our pipeline immediately reproducible, but we hope to see other implementations of the pipeline come about, perhaps using more niche robotics simulators or some of the popular, yet paid traffic simulators. Overall, we are extremely excited to provide this contribution to the robotics field, and we hope it can bring the immense benefits of HPC computing to future projects.  \clearpage

    %
    % The appendices are optional.  This is the format for two or more.
    % If you do not wish to include an appendix, comment out these lines.
    % If you want just one, see the formatting guidelines.
    %
    \begin{appendices}
        \begin{subappendices}
            
    \section{Pipeline Usage Walkthrough}

\begin{itemize}
    \item The customized Docker image of Webots loaded with pip and several popular python libraries is available for download at URL1. 
    \item The customized Singularity image of Webots, resultant from a conversion of the above Docker image, is available for download at URL2. 
    \item Our initial release of the project pipeline (sans the above images, due to git file size constraints), is available for viewing and download at \url{https://github.com/clemsonbds/CAVSITAW}. This repository also contains a detailed walkthrough for deploying and working with the simulation pipeline. 
\end{itemize}

 \clearpage

    \section{PBS Job Script}
\begin{lstlisting}[language=bash]
#!/bin/bash
#PBS -N webots
#PBS -l select=1:ncpus=5:mem=93gb:interconnect=hdr,walltime=00:45:00
#PBS -J 1-48
#PBS -q dicelab

echo Generating new random routes...
singularity exec -B $TMPDIR:$TMPDIR webots_sumo.sif /usr/local/webots/projects/default/resources/sumo/bin/duarouter --route-files .../SIM_$(($PBS_ARRAY_INDEX % 8))_net/sumo.flow.xml --net-file .../SIM_$(($PBS_ARRAY_INDEX % 8))_net/sumo.net.xml --output-file .../SIM_$(($PBS_ARRAY_INDEX % 8))_net/sumo.rou.xml --randomize-flows true --seed $RANDOM

echo Starting Webots on 'hostname'
singularity exec -B $TMPDIR:$TMPDIR webots_sumo.sif xvfb-run -a webots --stdout --stderr --batch --mode=realtime .../SIM_$(($PBS_ARRAY_INDEX % 8)).wbt
\end{lstlisting} \clearpage

            %\inputfile{appendixC.tex}
        \end{subappendices}
    \end{appendices}

    \singlespacing                             % Single space the Bibliography

    %
    % The bibliography style.  Set this to whatever matches you discipline.
    % For example, Computer Science would likely use 'plain'.  You might
    % also want to change the name from '' to 'References'
    % or 'Work Cited'.
    %
    % 'plain'   gets you numbered references and citations (e.g., [1] Dyson).
    %
    % 'alpha'   gets you labels formed from an abbreviation of the authors'
    %           names and the year of publication.  If there is more than
    %           one author, it will use the first letter of up to the first
    %           three authors' last names.
    %
    %           Some examples:
    %               [DED01] F.W. Dyson, A.G. Edgar, and D.B. Denny ... 2001
    %               [DE01] F.W. Dyson, A.G. Edgar ... 2001
    %               [Dys01] F.W. Dyson ... 2001
    %
    % 'apalike' gets you labels formed from the authors' names and year of
    %           publication.
    %
    %           Some examples:
    %               [Dyson et al., 2001] F.W. Dyson, A.G. Edgar, and
    %                 D.B. Denny ... 2001
    %               [Dyson and Edgar, 2001] F.W. Dyson, A.G. Edgar ... 2001
    %               [Dyson, 2001] F.W. Dyson ... 2001
    %
    \addtotoc{Bibliography}{\printbibliography}
\end{document}